\documentclass[prl,twocolumn,nofootinbib]{revtex4}
\usepackage{amsfonts}
\usepackage[mathscr]{eucal}

\newcommand{\be}{\begin{equation}}\newcommand{\ee}{\end{equation}}
\newcommand{\bea}{\begin{eqnarray}}
\newcommand{\eea}{\end{eqnarray}}

\newcommand{\p}[1]{(\ref{#1})}

\begin{document}

\vspace*{1.5cm}
\begin{flushright}
ITP-UH-24/08\\
JINR-E2-2008-191
\end{flushright}

\title{Supersymmetric Calogero models by gauging}

\author{Sergey  Fedoruk$\,{}^{1}$,  Evgeny Ivanov$\,{}^{1}$, Olaf
Lechtenfeld$\,{}^{2}$}

\affiliation{\vspace{0.2cm} ${}^1$Bogoliubov Laboratory of  Theoretical Physics, JINR,
141980 Dubna, Moscow Region, Russia \\
${}^2$Institut f{\" u}r Theoretische Physik, Leibniz Universit{\" a}t Hannover,
Appelstra{\ss}e 2, D-30167 Hannover, Germany}

\begin{abstract}
\vspace{0.2cm} \noindent New superconformal extensions of $d{=}1$ Calogero-type systems are
obtained by gauging the ${\rm U}(n)$ isometry of matrix superfield models. We consider the
cases of ${\cal N}{=}1$, ${\cal N}{=}2$ and ${\cal N}{=}4$ one-dimensional supersymmetries.
The bosonic core of the ${\cal N}{=}1$ and ${\cal N}{=}2$ models is the standard conformal
$A_{n-1}$ Calogero system, whereas the ${\cal N}{=}4$ model is an extension of the ${\rm
U}(2)$--spin Calogero system.

\end{abstract}

\maketitle

\setcounter{footnote}{0}

\noindent{\it 1.~Introduction.} Superconformal extensions of Calogero model \cite{C}
provide nice examples of integrable supersymmetric quantum-mechanical systems and as such
are of vast interest from various points of view (see \cite{OP,Poly06} for the survey of
physical applications of Calogero model). In particular, by a conjecture of Gibbons and
Townsend \cite{GT}, ${\cal N}{=}4$ superconformal models might be closely related to
M-theory. While the ${\cal N}{=}2$ super Calogero models for any number of interacting
particles were constructed in full generality rather long ago \cite{FM} (see also
\cite{Wyl1,BGK}), until now there is no such an exhaustive understanding of the generic
${\cal N}{=}4$ models despite the existence of extensive literature on this subject (see
e.g. \cite{Wyl1,BGK,BGL,GLP,KLP,BKS}). It seems important to develop some universal
approach to superconformal Calogero-type models including the ${\cal N}{=}4$ ones.

The purpose of this letter is to present a candidate approach of this type suitable for an
arbitrary number of interacting particles. It is based on the superfield gauging  of some
non-abelian isometries of the $d{=}1$ field theories. This gauging procedure was worked out
in \cite{DI} to understand off--shell dualities between $d{=}1$ supermultiplets with
different sets of physical bosonic components.

Our starting point is the nice interpretation of the bosonic $n$-particle Calogero model as
a ${\rm U}(n)$, $d{=}1$ gauge theory \cite{Poly91} (see also \cite{Gorsky} and
\cite{Poly011,Poly01}). In the formulation of \cite{Poly91} the model is described by the
hermitian $n{\times}n$--matrix field $ X_a^b(t)$, $(\overline{X_a^b}) =X_b^a $, the complex
$n$-plet $ Z_a(t)$, $\bar Z^a = (\overline{Z_a})$, $a,b=1,\ldots ,n$, and $n^2$
non--propagating ``gauge fields'' $ A_a^b(t)$, $(\overline{A_a^b}) =A_b^a$. The action
reads
\begin{equation}\label{b-Cal}
S_0 = \int dt  \,\Big[\, {\rm Tr}\left(\nabla\! X \nabla\! X \right) +
{\textstyle\frac{i}{2}} (\bar Z \nabla\! Z - \nabla\! \bar Z Z) + c\,{\rm Tr} A \,\Big],
\end{equation}
where the covariant derivatives are defined as
$$
\nabla\! X = \dot X +i [A,X], \quad \nabla\! Z = \dot Z + iAZ.
$$
The real constant $c$ of the Calogero interaction comes out from a Fayet-Iliopoulos (FI)
term in \p{b-Cal}.

The action~(\ref{b-Cal}) is invariant with respect to the local ${\rm U}(n)$
transformations, $g(\tau )\in {\rm U}(n)$,
\begin{equation}\label{ga-tr}
X \!\rightarrow  g X g^\dagger , \quad  Z \!\rightarrow  g Z , \quad A \!\rightarrow
 g A g^\dagger +i
\dot g g^\dagger,
\end{equation}
and we can fully fix the ${\rm U}(n)$ gauge freedom by choosing
\begin{equation}\label{ga-fi}
X_a^b = x_a \delta_a^b , \qquad  \bar Z^a = Z_a.
\end{equation}
Inserting these gauge conditions and the {\it algebraic} equations of motion $(Z_a)^2 =c$
(which implies $c>0$) and $A_a^b = \frac{Z_a Z_b}{2(x_a - x_b)^2}$, $a\neq b\,,$ back into
the action~(\ref{b-Cal}), we arrive at the standard Calogero action
\begin{equation}\label{st-Cal}
S_C = \int dt  \,\Big[\, \sum_{a} \dot x_a \dot x_a - \sum_{a\neq b} \frac{c^2}{4(x_a -
x_b)^2}\,\Big]\,
\end{equation}
as a fixed gauge of \p{b-Cal}. Note the important role of the auxiliary ${\rm U}(n)$
multiplet $Z$ with the $d{=}1$ Wess-Zumino (WZ) action in \p{b-Cal} for recovering the
Calogero action.

The original action \p{b-Cal} is invariant  under the $d{=}1$ conformal ${\rm SO}(1,2)$
transformations: $\delta t = a$, $ \delta X_a^b = {\textstyle\frac{1}{2}}\, \dot{a} X_a^b$,
$\delta Z_a = 0$, $\delta A_a^b = -\dot{a}A_a^b$, where $a(t)$ obeys the constraint
$\dddot{a} = 0 \,$. This property implies the well-known conformal invariance of the
eventual Calogero model.

Our approach is a minimal superfield generalization of this bosonic ${\rm U}(n)$ gauging.
Requiring the supersymmetric gauge models to possess ${\cal N}$--extended superconformal
symmetry essentially constrains the structure of the corresponding actions and allows one
to reveal, in their bosonic sector, either the standard Calogero model \p{st-Cal} (in the
cases of ${\cal N}{=}1$ and ${\cal N}{=}2$) or the U(2)--spin Calogero
model~\cite{GH-W,Poly02,Poly06} modified by a conformal potential for the center-of-mass
coordinate (in the ${\cal N}{=}4$ case). In this short note we outline the basic features
of our construction, leaving details, quantization and comparison with the previously known
superextended Calogero models for a longer paper.

\medskip

\noindent{\it 2.~${\cal N}{=}1$ supersymmetric extension$\,$}. We use the Grassmann-even
hermitian ${\cal N}{=}1$ matrix superfield $\mathscr{X}_a^b(t, \theta)$,
$(\mathscr{X})^\dagger =\mathscr{X}\,$, belonging to the adjoint representation of ${\rm
U}(n)$, as well as the Grassmann-even complex ${\cal N}{=}1$  superfield $ \mathcal{Z}_a
(t, \theta)$, $\bar \mathcal{Z}^a (t, \theta) = (\mathcal{Z}_a)^\dagger \,$, in the
fundamental of  ${\rm U}(n)\,$. The spinor and time derivatives,
$$
D = \partial_{\theta} +i\theta\partial_{t}\,, \quad \quad \{D, D \} = 2i\,
\partial_{t}\,,
$$
are gauge-covariantized by the anti--hermitian Grassmann-odd connections $
\mathscr{A}_a^b(t, \theta )$, $(\mathscr{A})^\dagger = -\mathscr{A}$:
$$
\mathscr{D} \mathscr{X} =  D \mathscr{X} +i [ \mathscr{A} , \mathscr{X}], \,\, \nabla_t
\mathscr{X} = -i\mathscr{D}\mathscr{D} \mathscr{X}, \,\, \mathscr{D} \mathcal{Z} =  D
\mathcal{Z} +i \mathscr{A} \mathcal{Z}\, .
$$
The minimal gauge invariant action has the following form ($\mu_1=dt d\theta$):
\begin{equation}\label{N1-Cal}
S_{1}\! =\! -i\!\int\! \mu_1 \Big[ {\rm Tr} \left( \nabla_t \mathscr{X} \mathscr{D}
\mathscr{X} + c \mathscr{A} \right) + {\textstyle\frac{i}{2}}(\bar \mathcal{Z} \mathscr{D}
\mathcal{Z} - \mathscr{D}\bar \mathcal{Z} \mathcal{Z} )\Big].
\end{equation}
It is invariant under the local ${\rm U}(n)$ transformations: \bea \mathscr{X}^{\,\prime} =
e^{i\tau} \mathscr{X} e^{-i\tau}\! , \,\mathcal{Z}^{\prime} = e^{i\tau} \mathcal{Z} , \,
\mathscr{A}^{\,\prime} =  e^{i\tau} \mathscr{A} e^{-i\tau} - i e^{i\tau} D e^{-i\tau}\,,
\nonumber \eea where $ \tau_a^b(t, \theta) \in u(n) $ is the hermitian matrix parameter. It
is also invariant under the ${\cal N}{=}1$ superconformal group. Conformal supersymmetry
$\delta^\prime t= -i\,\eta\theta t$, $ \delta^\prime \theta = \eta t$ has the following
realization on the involved superfields:
$$
\delta^\prime \mathscr{X}= -i\,\eta\theta\,\mathscr{X},\quad \delta^\prime \mathscr{A} =
i\,\eta\theta\,\mathscr{A},\quad \delta^\prime \mathcal{Z} = 0.
$$
Its closure with the Poincar\'e ${\cal N}{=}1$, $d{=}1$ supersymmetry yields the full
${\cal N}{=}1$ superconformal symmetry.

Due to the ${\rm U}(n)$ gauge invariance we can choose the WZ gauge for the spinor
connection:
\begin{equation}\label{1N-WZ}
\mathscr{A} = i\theta A(t)\,.
\end{equation}
Substituting this into the action~(\ref{N1-Cal}), integrating there over $\theta$ and
eliminating the auxiliary fields by their equations of motion, we obtain
\begin{equation}\label{1N-Cal-WZ}
S_{1} = S_{0}+S^\Psi_{1},\qquad S^\Psi_{1}=-i\,{\rm Tr}\int dt \,\Psi \nabla \Psi
\end{equation}
where $\Psi = -iD\mathscr{X}|$ is the matrix Grassmann-odd field and $\nabla \Psi = \dot
\Psi +i [A,\Psi]$. The bosonic limit of (\ref{1N-Cal-WZ}) (and hence of (\ref{N1-Cal})) is
just the Calogero action~(\ref{b-Cal}). Its gauge ${\rm U}(n)$ symmetry is the residual
symmetry of the WZ gauge \p{1N-WZ}.

An alternative supersymmetric gauge choice is
\begin{equation}\label{s-X-fix1}
\mathscr{X}_a^b =0,\quad a\neq b \quad \Leftrightarrow \quad \mathscr{X}_a^b =
\mathscr{X}_a\delta^b_a\,;\qquad \mathcal{Z}_a = \bar\mathcal{Z}^{a}.
\end{equation}
In this gauge the model is described by $n^2$ real ${\cal N}{=}1$ superfields
$\mathscr{A}_a^b, a\neq b\,,$ and $\mathscr{X}_a\,$ (the superfields $\mathcal{Z}_a$ and
those on the diagonal of $\mathscr{A}_a^b$ are auxiliary). In the two-particle case
($n{=}2$) the resulting ${\cal N}{=}1$ system possesses an additional hidden ${\cal N}{=}1$
supersymmetry, so that the $n{=}2$ model is in fact ${\cal N}{=}2$ superconformal mechanics
plus an ${\cal N}{=}2$ free multiplet corresponding to the center-of-mass motion. Starting
with $n{=}3\,$, one gets new ${\cal N}{=}1$ superextensions of the $n$-particle Calogero
models which cannot be recovered by any truncations of the standard ${\cal N}{=}2$
superextensions \cite{FM}.

\medskip

\noindent{\it 3.~${\cal N}{=}2$ supersymmetric extension$\,$}. The relevant superfield
content consists of the hermitian matrix superfield $ \mathscr{X}_a^b(t,
\theta,\bar\theta)$, $(\mathscr{X})^\dagger =\mathscr{X}$ with the off-shell field content
${\bf (1, 2, 1)}$, and bosonic chiral ${\rm U}(n)$--fundamental superfield $ \mathcal{Z}_a
(t_{\!\scriptscriptstyle{L}}, \theta)$, $\bar \mathcal{Z}^a (t_{\!\scriptscriptstyle{R}},
\bar\theta) = (\mathcal{Z}_a)^\dagger$, $t_{\!\scriptscriptstyle{L,R}}=t\pm
i\theta\bar\theta$,
\begin{equation}\label{N2-chir-Z}
\bar D \mathcal{Z}_a=0 \,, \qquad D \bar \mathcal{Z}^a=0\,,
\end{equation}
with the field contents ${\bf (2,2,0)}$. Here
$$
D = \partial_{\theta} +i\bar\theta\partial_{t}\,, \quad \bar D = -\partial_{\bar\theta}
-i\theta\partial_{t}\,, \quad \{D, \bar D \} = -2i \partial_{t}\,.
$$
The gauge prepotential is an $n{\times}n$ hermitian matrix $ V_a^b(t, \theta,\bar\theta)$,
$(V)^\dagger = V$. The action reads ($\mu_2=dt d^2\theta $)
\begin{equation}\label{N2-Cal}
S_{2} \!=\! \int \!\mu_2 \Big[ {\rm Tr}\! \left( \bar\mathscr{D} \mathscr{X} \, e^{2V}
\mathscr{D} \mathscr{X}\, e^{2V} \right) + {\textstyle\frac{1}{2}}\bar \mathcal{Z}\,
e^{2V}\! \mathcal{Z} - c\,{\rm Tr} V  \Big],
\end{equation}
the gauge-covariant derivatives being defined as
$$
\mathcal{D} \mathscr{X} =  D \mathscr{X} + e^{-2V} (D e^{2V}) \, \mathscr{X}, \,
\bar\mathcal{D} \mathscr{X} = \bar D \mathscr{X} - \mathscr{X} \, e^{2V} (\bar D e^{-2V}).
$$
The action~(\ref{N2-Cal}) is invariant under the local ${\rm U}(n)$ transformations:
\begin{equation}\label{tran-X}
\mathscr{X}^{\,\prime} =  e^{i\lambda} \mathscr{X} e^{-i\bar\lambda}, \quad
\mathcal{Z}^{\prime} =  e^{i\lambda} \mathcal{Z} , \quad e^{2V^\prime} = e^{i\bar\lambda}
  e^{2V} e^{-i\lambda}
\end{equation}
where $n^2$ complex gauge parameters, $\lambda= (\lambda_a^b)\,$, are (anti)chiral
superfields: $ \lambda (t_{\!\scriptscriptstyle{L}}, \theta) \in u(n) $, $\bar\lambda
(t_{\!\scriptscriptstyle{R}}, \theta) = (\lambda)^\dagger  \in u(n) $. The action is also
invariant under the superconformal group ${\rm SU}(1,1|1)\,$. The conformal supersymmetry
acts on the coordinates as
$$
\delta^\prime t= -i(\eta\bar\theta+\bar\eta\theta)t, \quad\delta^\prime \theta =
\eta(t+i\theta\bar\theta)
$$
and on the superfields as
$$
\delta^\prime \mathscr{X}= -i(\eta\bar\theta+\bar\eta\theta)\,\mathscr{X}, \quad
\delta^\prime V = 0,\quad \delta^\prime \mathcal{Z} = 0 .
$$
The chirality conditions~(\ref{N2-chir-Z}) are preserved by these transformations.

In the WZ gauge
\begin{equation}\label{WZ-2}
V (t, \theta,\bar\theta) = -\theta\bar\theta A (t)
\end{equation}
the action~(\ref{N2-Cal}) takes the form
\begin{equation}\label{N2Cal-com}
S_{2} = S_{0}+ S^{\Psi}_{2}, \quad S^{\Psi}_{2}= -i\, {\rm Tr} {\displaystyle\int} dt \,(
\bar\Psi \nabla \Psi - \nabla \bar\Psi \Psi )
\end{equation}
where $\Psi = D \mathscr{X}|$ is a Grassmann-odd field and
\begin{equation}\label{cov-der-Psi}
\nabla \Psi = \dot \Psi +i [A,\Psi]\,, \qquad \nabla \bar\Psi = \dot {\bar\Psi} +i
[A,\bar\Psi]\,.
\end{equation}
We see that the bosonic core of the action~(\ref{N2Cal-com}) exactly coincides with the
Calogero action~(\ref{b-Cal}).

The action~(\ref{N2Cal-com}) is invariant with respect to the residual local bosonic ${\rm
U}(n)$ transformations, defined by~(\ref{ga-tr}) and $\Psi \rightarrow g \Psi g^\dagger$,
therefore we can choose the gauge~(\ref{ga-fi}). As a result we obtain an ${\cal N}{=}2$
superextension of the $n$-particle Calogero model. In the two-particle case ($n{=}2$) we
found that the ${\cal N}{=}2$ supersymmetric gauged system actually describes ${\cal
N}{=}4$ superconformal mechanics plus one ${\cal N}{=}4$ free multiplet corresponding to
the center-of-mass motion, so that there is a hidden extra ${\cal N}{=}2$ symmetry in this
case. For $n> 2$ we obtain some new ${\cal N}{=}2$ extensions of the $n$-particle Calogero
models with $n$ bosonic variables and $n{\times} n$ fermionic ones, as opposed to the
standard ${\cal N}{=}2$ super-Calogero with $n$ complex fermions \cite{FM}.

The presence of the matrix Grassmann-odd field $\Psi$ in the action~(\ref{N2Cal-com}) (and
also in~(\ref{1N-Cal-WZ})) is imperative for $d{=}1$ supersymmetry and superconformal
symmetry. Similar structures with the bosonic analogs of $\Psi$ appeared e.g. in
\cite{Poly011,Poly01} in connection with the quantum Hall effect.

\medskip

\noindent{\it 4.~${\cal N}{=}4$ supersymmetric extension.} This case surprisingly yields
${\rm U}(2)$-spin Calogero system \cite{GH-W,Poly02} in the bosonic sector. The most
natural formulation of ${\cal N}{=}4, d{=}1$ models is achieved in the harmonic superspace
\cite{GIOS,IL} parametrized by the coordinates $(t,\theta_i, \bar\theta^k, u_i^\pm)$, $i,k
=1,2$, where commuting mutually conjugate ${\rm SU}(2)$--doublets $u_i^\pm$ are harmonic
coordinates, $u^{+i}u_i^-=1$. The harmonic analytic subspace is parametrized by the
coordinates $(\zeta,u)=(t_A,\theta^+, \bar\theta^+, u_i^\pm)$, $t_A=t-i(\theta^+
\bar\theta^- +\theta^-\bar\theta^+)$, $\theta^\pm=\theta^i u_i^\pm$,
$\bar\theta^\pm=\bar\theta^i u_i^\pm\,$. The integration measures are defined as $\mu_H
=dudtd^4\theta$ and $\mu^{(-2)}_A=dud\zeta^{(-2)}$.

The ${\cal N}{=}4$ supersymmetric model with ${\rm U}(n)$ gauge symmetry is described by
the action
\begin{equation}\label{4N-gau}
S_4 =S_{\mathscr{X}} + S_{FI} + S_{WZ}\,.
\end{equation}

The first term in~(\ref{4N-gau})
\begin{equation}\label{4N-X}
S_{\mathscr{X}} =-{\textstyle\frac{1}{2}}\int \mu_H  {\rm Tr} \left( \mathscr{X}^{\,2} \,
\right)
\end{equation}
is the gauged action of the {\bf (1,4,3)} multiplets. The latter are described by hermitian
matrix superfields $\mathscr{X}=(\mathscr{X}_a^b)$ subjected to the gauge-covariant
constraints
\begin{eqnarray}
&&\mathscr{D}^{++} \,\mathscr{X}=0, \label{cons-X-g-V}\\
\mathscr{D}^{+}\mathscr{D}^{-} \,\mathscr{X}=0,&&
  (\mathscr{D}^{+}\bar\mathscr{D}^{-} +\bar\mathscr{D}^{+}\mathscr{D}^{-})\,
\mathscr{X}=0.
  \label{cons-X-g}
\end{eqnarray}
The constraint~(\ref{cons-X-g-V}) involves the covariant harmonic derivative
$\mathscr{D}^{++} = D^{++} + i\,V^{++}$, where the gauge matrix connection
$V^{++}(\zeta,u)$ is an analytic superfield.\footnote{Besides the covariant derivative
$\mathscr{D}^{++}$ which commutes with $D^+, \bar D^+$ and so preserves the analyticity,
one can define the derivative $\mathscr{D}^{--} = D^{--} + i\,V^{--}$, so that
$[\mathscr{D}^{++}, \mathscr{D}^{--}] = D^0\,$ and $D^0$ is the operator counting the
external U(1) charges of superfields. The non-analytic connection $V^{--}$ is expressed
through $V^{++}$ from this commutation relation \cite{GIOS}.} The gauge connections
entering the spinor covariant derivatives in~(\ref{cons-X-g}) are properly expressed
through $V^{++}(\zeta,u)$ \cite{DI}. The parameters of the ${\rm U}(n)$ gauge group are
analytic, so $\mathscr{D}^{+} = D^{+}\,,\;\bar\mathscr{D}^{+} = \bar D^{+}$. Note that
$\mathscr{X}$ is in the adjoint of ${\rm U}(n)$, so $ \mathscr{D}^{++} \mathscr{X} = D^{++}
\mathscr{X} + i\,[V^{++} ,\mathscr{X}] $, etc.

The second term in~(\ref{4N-gau}) is the FI term
\begin{equation}\label{4N-FI}
S_{FI} ={\textstyle\frac{i}{2}}\,c\int \mu^{(-2)}_A  \,{\rm Tr} \,V^{++}  .
\end{equation}

The third term in~(\ref{4N-gau}),
\begin{equation}\label{4N-VZ}
S_{WZ} = {\textstyle\frac{1}{2}}\int \mu^{(-2)}_A  \mathcal{V}_0
\widetilde{\,\mathcal{Z}}{}^+ \mathcal{Z}^+\,,
\end{equation}
is a WZ action describing $n$ commuting analytic superfields $\mathcal{Z}^+_a$ (analogs of
the superfields $\mathcal{Z}_a$ of the ${\cal N}{=}1$ and ${\cal N}{=}2$ cases). They
represent off-shell ${\cal N}{=}4$ multiplets {\bf (4,4,0)} and are defined by the
constraints
\begin{equation}  \label{cons-Ph-g}
\mathscr{D}^{++} \mathcal{Z}^+=0, \quad {D}^{+} \mathcal{Z}^+ =0\,,\quad
  \bar{D}^{+} \mathcal{Z}^+ =0\,.
\end{equation}

At last, the superfield $\mathcal{V}_0(\zeta,u)$ is a real analytic gauge superfield,
${D}^{+} \mathcal{V}_0=0$, $ \bar{D}^{+} \mathcal{V}_0=0$, which is defined by the integral
transform \cite{DI}
$$
\mathscr{X}_0(t,\theta_i,\bar\theta^i)=\int du \mathcal{V}_0 \left(t_A, \theta^+,
\bar\theta^+, u^\pm \right) \Big|_{\theta^\pm=\theta^i u^\pm_i,\,\,\,
\bar\theta^\pm=\bar\theta^i u^\pm_i}
$$
which resolves the constraints~\p{cons-X-g-V}, (\ref{cons-X-g}) for the singlet ${\rm
U}(1)$ part $\mathscr{X}_0 \equiv {\rm Tr} \left( \mathscr{X} \right)$.

The action~(\ref{4N-gau}) is invariant under the ${\cal N}{=}4$ superconformal group
$D(2,1;\alpha)$ with $\alpha = -\frac12$. To show this we should use the $D(2,1,\alpha)$
transformation laws given in \cite{IL,DI}, in particular that of conformal supersymmetry,
$$
\delta^\prime \mu_H=
\mu_H\left(2\Lambda-{\textstyle\frac{1+\alpha}{\alpha}}\,\Lambda_0\right),
\quad\delta^\prime \mu^{(-2)}_A= 0
$$
with $ \Lambda = 2i\alpha(\bar\eta^-\theta^+ - \eta^-\bar\theta^+)$, $\Lambda_0 = 2\Lambda-
D^{--} \Lambda^{++}$, $\Lambda^{++} = D^{++}\Lambda$. The involved $d{=}1$ superfields are
transformed as follows:
$$
\delta^\prime \mathscr{X}= -\Lambda_0\,\mathscr{X},\quad \delta^\prime \mathcal{Z}^+ =
\Lambda\,\mathcal{Z}^+,\quad \delta^\prime V^{++} = 0.
$$
The variation of the action~(\ref{4N-X}) is vanishing only at $\alpha=-\frac{1}{2}$,
whereas the constraints~(\ref{cons-X-g-V}), (\ref{cons-X-g}), (\ref{cons-Ph-g}), as well as
the actions~(\ref{4N-FI}), (\ref{4N-VZ}), are superconformally invariant for an arbitrary
parameter $\alpha$. It is important that just the field multiplier $\mathcal{V}_0$ in the
action~(\ref{4N-VZ}) provides this invariance due to its transformation law
$\delta'\mathcal{V}_0 = -2\Lambda \mathcal{V}_0$ \cite{DI}. Note that at $\alpha = -1/2$
the supergroup $D(2,1;\alpha)$ is isomorphic to OSp$(4|2)$ \cite{Sorba,Proe}, so our gauge
approach in the ${\cal N}{=}4$ case implies a different ${\cal N}{=}4$ superconformal group
as compared to the more customary SU$(1,1|2)$ used e.g. in \cite{GLP}.

The local ${\rm U}(n)$ transformations leaving the action~(\ref{4N-gau}) invariant are
given by
\begin{equation}\label{tran4}
\begin{array}{rrl}
&& \mathscr{X}^{\,\prime} =  e^{i\lambda} \mathscr{X} e^{-i\lambda} , \qquad
\mathcal{Z}^+{}^{\prime}
= e^{i\lambda} \mathcal{Z}^+ , \\
&& V^{++}{}^{\,\prime} =  e^{i\lambda}\, V^{++}\, e^{-i\lambda} - i\, e^{i\lambda} (D^{++}
e^{-i\lambda}),
\end{array}
\end{equation}
where $ \lambda_a^b(\zeta, u^\pm) \in u(n) $ is the ``hermitian'' analytic matrix
parameter, $\widetilde{\lambda} =\lambda$. Using this gauge freedom we can choose the WZ
gauge
\begin{equation}  \label{WZ-4N}
V^{++} =-2i\,\theta^{+}
  \bar\theta^{+}A(t_A) .
\end{equation}
In this gauge we have
\begin{equation}\nonumber
\begin{array}{rrl}
&& \mathscr{D}^{\pm\pm}= D^{\pm\pm} +2\,\theta^{\pm}
  \bar\theta^{\pm}\,A, \\
&& \mathscr{D}^{-}=D^{-} -2\,
  \bar\theta^{-}\,A, \quad \bar\mathscr{D}^{-}= \bar D^{-} -2\,
  \theta^{-}\,A
\end{array}
\end{equation}
and the constraints~(\ref{cons-X-g-V}), (\ref{cons-X-g}) are solved by
\begin{equation}  \label{X-WZ}
\mathscr{X}= X + \theta^-\bar\theta^- N^{++} + \theta^- \Psi^+ + \bar\theta^- \bar\Psi^+ +
\ldots \,,
\end{equation}
where $N^{++} = N^{ik}u_i^+ u_k^+ $, $\Psi^{+} = \Psi^{i}u_i^+ $, $\bar\Psi^{+} =
\bar\Psi^{i}u_i^+$ and the fields $X(t_A)$, $N^{ik}= N^{(ik)}(t_A)$, $\Psi^{i}(t_A)$ ,
$\bar\Psi^{i}(t_A)$ are ordinary $d{=}1$ fields having no dependence of the harmonics. All
other fields in~(\ref{X-WZ}) are expressed through these fields and their covariant
derivatives $\nabla_{t_A}X =
\partial_{t_A}X +i[ A,X]$, etc. The solution of the constraints~(\ref{cons-Ph-g}) is
\begin{equation}  \label{Ph-WZ}
\mathcal{Z}^+ = Z^+ + \theta^+ \varphi + \bar\theta^+ \phi + 2i\, \theta^+ \bar\theta^+
Z^-,
\end{equation}
where $Z^{+} = Z^{i}(t_A)u_i^+  $, $Z^{-} = \nabla_{t_A}Z^{i}(t_A)u_i^-$.

Inserting the expressions~(\ref{X-WZ}), (\ref{Ph-WZ}) in the action~(\ref{4N-gau}) and
eliminating the fields $N^{ik}$, $\phi$,  $\bar\phi$, $\varphi$,  $\bar\varphi$ by their
equations of motion we obtain, in the WZ gauge,
\begin{eqnarray}\label{4N-gau-bose-a}
S_4  &=& S_b + S_f,
\\
S_b &=&  \int dt \,\Big[{\rm Tr} \left( \nabla X\nabla X +c \,A \right)
+ {\textstyle\frac{n}{8}}(\bar Z^{(i} Z^{k)})(\bar Z_{i} Z_{k})\nonumber \\
&& + {\textstyle\frac{i}{2}}\,X_0 \left(\bar Z_k \nabla Z^k - \nabla \bar Z_k \, Z^k\right)
\Big],\label{4N-gau-bose-1}
\\
  S_f&=&   -i{\rm Tr} \int dt \left( \bar\Psi_k \nabla\Psi^k
-\nabla\bar\Psi_k \Psi^k
\right) \nonumber \\
&& -\int dt  \,\frac{\Psi^{(i}_0\bar\Psi^{k)}_0 (\bar Z_{i}
Z_{k})}{X_0}\,,\label{4N-gau-fermi-1}
\end{eqnarray}
where
$$
X_0 \equiv {\rm Tr} (X), \quad\Psi_0^i \equiv {\rm Tr} (\Psi^i), \quad\bar\Psi_0^i \equiv
{\rm Tr} (\bar\Psi^i).
$$

Let us consider the bosonic limit of $S_4$, i.e. the action~(\ref{4N-gau-bose-1}). We can
impose the gauge $X_a^b =0$, $a\neq b$, using the residual invariance of WZ
gauge~(\ref{WZ-4N}): $ X^{\,\prime} =  e^{i\lambda}\, X\, e^{-i\lambda} $,
$Z^{\prime}{}^{k} =  e^{i\lambda} Z^{k}$, $ A^{\,\prime} =  e^{i\lambda}\, A\,
e^{-i\lambda} - i\, e^{i\lambda} (\partial_t e^{-i\lambda})$ where $ \lambda_a^b(t) \in
u(n) $ are ordinary $d{=}1$  gauge parameters. As a result of this, and after eliminating
$A_a^b$, $a\neq b$, by the equations of motion, the action~(\ref{4N-gau-bose-1}) takes the
following form (instead of $Z^i_a$ we introduce the new fields $ Z^\prime{}^i_a =
(X_0)^{1/2}\,Z^i_a$ and omit the primes on these fields),
\begin{eqnarray}
S_{b} &=& \int dt \Big\{ \sum_{a} \dot x_a \dot x_a + {\textstyle\frac{i}{2}}\sum_{a} (\bar
Z_k^a \dot Z^k_a - \dot {\bar Z}{}_k^a Z^k_a)
+ \nonumber\\
&& \quad  + \sum_{a\neq b} \, \frac{{\rm Tr}(S_a S_b)}{4(x_a - x_b)^2} - \frac{n\,{\rm
Tr}(\hat S \hat S)}{2(X_0)^2}\,\Big\}. \label{4N-bose-fix}
\end{eqnarray}
Here, the fields $Z^k_a$ are subject to the constraints \footnote{Here and in~(\ref{S}) we
do not sum over the repeated index $a$.}
\begin{equation}\label{4N-eq-aa}
\bar Z_i^a Z^i_a =c \qquad \forall \, a \,,
\end{equation}
and carry the residual abelian gauge $[{\rm U}(1)]^n$ symmetry, $Z_a^k \rightarrow
e^{i\varphi_a} Z_a^k\,$, with local parameters $\varphi_a(t)$. In~(\ref{4N-bose-fix}) we
use the following notation:
\begin{eqnarray}\label{S}
(S_a)_i{}^j &\equiv& \bar Z^a_i Z_a^j,\\
(\hat S)_i{}^j &\equiv& \sum_a \left[ (S_a)_i{}^j -
{\textstyle\frac{1}{2}}\delta_i^j(S_a)_k{}^k\right].\label{hS}
\end{eqnarray}
Note that at $c=0$ the constraint \p{4N-eq-aa} implies $Z^i_a =0$, i.e. a non-trivial
interaction exists only for $c\neq 0$ as in the previous cases. The new feature of the
${\cal N}{=}4$ case is that not all out of the bosonic variables $Z^i_a$ are eliminated by
fixing gauges and solving the constraint; there survives a non-vanishing WZ term for them
in eq. (\ref{4N-bose-fix}). After quantization these variables become purely internal
(${\rm U}(2)$-spin) degrees of freedom.

In the Hamiltonian approach, the kinetic WZ term for $Z$ in~(\ref{4N-bose-fix}) gives rise
to the following Dirac brackets:
\begin{equation}\label{DB}
[\bar Z^a_i, Z_b^j]_{{}_D}= i\delta^a_b\delta_i^j.
\end{equation}
With respect to these brackets the quantities~(\ref{S}) for each index $a$ form $u(2)$
algebras
\begin{equation}\label{su-DB}
[(S_a)_i{}^j, (S_b)_k{}^l]_{{}_D}= i\delta_{ab}\left\{\delta_i^l(S_a)_k{}^j-
\delta_k^j(S_a)_i{}^l \right\}.
\end{equation}
The quantities~(\ref{hS}) are time-independent Noether charges for the ${\rm SU}(2)$
invariance of the system~(\ref{4N-bose-fix}), so the numerator of the term $\sim
(X_0)^{-2}$ in \p{4N-bose-fix} is a constant on the equations of motion for $Z_a^i, \bar
Z^a_i\,$. So, as opposed to the ${\cal N}{=}1,2$ cases, the ${\cal N}{=}4$ action contains
a conformal potential even in the center-of-mass sector (like in \cite{GLP,KLP}). Modulo
this extra conformal potential (last term in~(\ref{4N-bose-fix})), the bosonic limit of the
${\cal N}{=}4$ system constructed is none other than the integrable U(2)--spin Calogero
model \cite{GH-W} in the formulation of \cite{Poly02,Poly06}.

While the coordinate $X_0$ decouples in the bosonic limit, when all fermions are discarded,
this is not the case for the full action because of the term $\sim X_0^{-1}$ in
\p{4N-gau-fermi-1}. The full SU(2) current contains extra fermionic terms, and its bosonic
part \p{hS} is not conserved by itself.

\medskip

\noindent{\it 5.~Outlook}. In this paper we proposed a new gauge approach to the
construction of superconformal Calogero-type systems as a superextension of the bosonic
construction of \cite{Poly91}. The characteristic features of this approach are the
presence of auxiliary supermultiplets with WZ type actions, the built-in superconformal
invariance and the emergence of the Calogero coupling constant as a strength of the FI term
of the U(1) gauge (super)field. Here we used the ${\rm U}(n)$ gauging and obtained new
superextensions of the $A_{n-1}$ Calogero model and of its U(2)-spin extension (in the
${\cal N}{=}4$ case). Superextensions of other conformal Calogero models could be
presumably obtained by choosing other gauge groups and/or representations for the matrix
and WZ superfields. Superextensions of non-conformal models can be constructed by adding
proper gauge invariant (but not conformally invariant) potentials to the original
superfield actions.

While in the ${\cal N}{=}1$ and ${\cal N}{=}2$ cases there is almost no freedom in the
choice of the original gauged action (provided that it is required to be minimal and
superconformal), it is not so in the ${\cal N}{=}4$ case due to the diversity of the ${\cal
N}{=}4, d{=}1$ multiplets. For instance, any sort of the ${\cal N}{=}4, d{=}1$ multiplet
has its ``mirror'' in which another SU(2) from the full R-symmetry group SO(4) of the
${\cal N}{=}4, d{=}1$ superalgebra is manifest. We are planning to consider these
possibilities elsewhere.

In the ${\cal N}{=}4$ case we used as the basic matrix superfield the new non-abelian
version of the multiplet ${\bf (1,4,3)}$ defined by the constraints ~(\ref{cons-X-g-V}) and
(\ref{cons-X-g}). Its simplest, quadratic action is invariant under the superconformal
group $D(2,1;\alpha)$ with $\alpha = -1/2\,$ (which is isomorphic to OSp$(4|2)$). It is
worth noting that our gauging procedure is compatible as well with other ${\cal N}{=}4$
superconformal groups~\cite{Sorba,Proe,IL,IKL}. For any value of $\alpha\neq 0$ the
superconformal ${\cal N}{=}4$ gauged action has the generic form of~(\ref{4N-gau}) with the
same $S_{WZ}$ and $S_{FI}$, the only difference being in the form of the action for
$\mathscr{X}$,
\begin{equation}\label{4N-X-a}
S_{\mathscr{X}}^{\alpha\neq 0}
  =\alpha\int \mu_H \,\Big[ {\rm Tr} \left( \mathscr{X}^{\,2}
\, \right)\Big]^{-{\textstyle\frac{1}{2\alpha}}}\, .
\end{equation}
It is important that $\mathscr{X}$ is a matrix and, therefore, this action is non-trivial
even in the case of $\alpha=-1$ as opposed to the standard case of the abelian ${\bf
(1,4,3)}$ multiplet. The second possibility at $\alpha=-1$ is to consider the matrix
version  of the standard conformal action
\begin{equation}\label{4N-X-1}
\tilde{S}_{\mathscr{X}}^{\alpha=-1}
  = \int \mu_H \, {\rm Tr} \Big( \mathscr{X}\,\ln\mathscr{X}
\, \Big).
\end{equation}
It seems, however, that all such actions except for the case of $\alpha = -1/2$ yield
non-trivial sigma-model type kinetic terms for the $X$ fields, so the corresponding bosonic
limits are some more general conformal models.

\smallskip

{\it Acknowledgements.} We thank Francois Delduc and Armen Nersessian for the interest in
this work. We acknowledge a support from a grant of the Heisenberg--Landau Programme, RFBR
grants 06-02-16684, 08-02-90490 and INTAS grant  05-1000008-7928 (S.F. \& E.I.) and a DFG
grant, project No 436 RUS/113/669 (E.I. \& O.L.).


\smallskip


\begin{thebibliography}{96}

\bibitem{C}
F.~Calogero, J. Math. Phys. {\bf 10}, 2191 (1969); {\bf 10}, 2197 (1969);  {\bf 12}, 419
(1971).

\bibitem{OP}
M.A.~Olshanetsky, A.M.~Perelomov, Phys. Rept. {\bf 71}, 313 (1981);  {\bf 94}, 313 (1983).

\bibitem{Poly06}
A.P.~Polychronakos,  J. Phys. {\bf A39}, 12793 (2006), {\tt arXiv:hep-th/0607033}.

\bibitem{GT}
G.W.~Gibbons, P.K.~Townsend Phys. Lett. {\bf B454}, 187 (1999), {\tt arXiv:hep-th/9812034}.

\bibitem{FM}
D.Z.~Freedman, P.F.~Mende, Nucl. Phys. {\bf B344}, 317 (1990).

\bibitem{Wyl1}
N.~Wyllard, J. Math. Phys. {\bf 41}, 2826 (2000), {\tt arXiv:hep-th/9910160}.

\bibitem{BGK}
S.~Bellucci, A.~Galajinsky, S.~Krivonos, Phys. Rev. {\bf D68}, 064010 (2003), {\tt
arXiv:hep-th/0304087}.

\bibitem{BGL}
S.~Bellucci, A.V.~Galajinsky, E.~Latini, Phys. Rev. {\bf D71}, 044023 (2005), {\tt
arXiv:hep-th/0411232}.

\bibitem{GLP}
A.~Galajinsky, O.~Lechtenfeld, K.~Polovnikov, Phys. Lett. {\bf B643}, 221 (2006), {\tt
arXiv:hep-th/0607215}; JHEP {\bf 0711}, 008 (2007), {\tt arXiv:0708.1075 [hep-th]}; {\it
N=4 mechanics, WDVV equations and roots}, {\tt arXiv:0802.4386 [hep-th]}.

\bibitem{BKS}
S.~Bellucci, S.~Krivonos, A.~Sutulin, Nucl. Phys. {\bf B805}, 24 (2008), {\tt
arXiv:0805.3480 [hep-th]}.

\bibitem{KLP}
S. Krivonos, O. Lechtenfeld, K. Polovnikov,
    {\it N=4 superconformal n-particle mechanics via superspace}, {\tt
arXiv:0812.5062 [hep-th]}.

\bibitem{DI}
F.~Delduc, E.~Ivanov, Nucl. Phys.  {\bf B753}, 211 (2006), {\tt arXiv:hep-th/0605211}; {\bf
B770}, 179 (2007), {\tt arXiv:hep-th/0611247}; {\bf B787}, 176 (2007), {\tt arXiv:0706.0706
[hep-th]}; Phys. Lett. {\bf B654}, 200 (2007), {\tt arXiv:0706.2472 [hep-th]}.

\bibitem{Poly91}
A.P.~Polychronakos,  Phys. Lett. {\bf B266}, 29 (1991).

\bibitem{Gorsky}
A.~Gorsky, N.~Nekrasov, Nucl. Phys. {\bf B414} 213 (1994), {\tt arXiv:hep-th/9304047};
ibid. {\bf B436} 582 (1995), {\tt arXiv:hep-th/9401017}; Theor. Math. Phys. {\bf 100} 874
(1994).

\bibitem{Poly011} A.P.~Polychronakos, JHEP {\bf 0104}, 011 (2001), {\tt
arXiv:hep-th/0103013}.

\bibitem{Poly01}
B. Morariu, A.P.~Polychronakos, JHEP {\bf 0107}, 006 (2001), {\tt arXiv:hep-th/0106072};
Phys. Rev. {\bf D72}, 125002 (2005), {\tt arXiv:hep-th/0510034}.

\bibitem{GH-W}
J.~Gibbons, T.~Hermsen, Physica {\bf D11}, 337 (1984); S.~Wojciechowski, Phys. Lett. {\bf
A111}, 101 (1985).

\bibitem{Poly02}
A.P.~Polychronakos, Nucl. Phys. {\bf B543}, 485 (1999), {\tt arXiv:hep-th/9810211}; Phys.
Rev. Lett. {\bf 89}, 126403 (2002), {\tt arXiv:hep-th/0112141}.

\bibitem{GIOS}
A.S.~Galperin, E.A.~Ivanov, V.I.~Ogievetsky, E.S.~Sokatchev, {\it Harmonic superspace},
Cambridge University Press, 2001, 306 p.

\bibitem{IL}
E.~Ivanov, O.~Lechtenfeld, JHEP {\bf 0309}, 073 (2003), {\tt arXiv:hep-th/0307111}.

\bibitem{IKL}
E.~Ivanov, S.~Krivonos, O.~Lechtenfeld, JHEP {\bf 0303}, 014 (2003), {\tt
arXiv:hep-th/0212303}.

\bibitem{Sorba}
L.~Frappat, A.~Sciarrino, P.~Sorba, {\it Dictionary on Lie Superalgebras}, {\tt
arXiv:hep-th/9607161}.

\bibitem{Proe}
A.~Van~Proeyen, {\it Tools for supersymmetry}, {\tt arXiv:hep-th/9910030}.

\end{thebibliography}
\end{document}